# Existence of the Co$^{3+}$ Low Spin State in TbBaCo$_2$O$_{5.5}$

Minoru Soda[1], Yukio Yasui[1], Yoshiaki Kobayashi[1], Toshiaki Fujita[1], Masatoshi Sato[1*] and Kazuhisa Kakurai[2]

[1]*Department of Physics, Division of Material Science, Nagoya University, Furo-cho, Chikusa-ku, Nagoya 464-8602*

[2] *Quantum Beam Science Directorate, Japan Atomic Energy Agency, Tokai-mura, Naka-gun, Ibaraki 319-1195*

**Abstract**

NMR studies have been carried out on a single crystal sample of TbBaCo$_2$O$_{5.5}$ with the oxygen-deficient perovskite structure. We have also carried out neutron diffraction studies and added new results to our previously published data. By analyzing these data, the magnetic structures and the spin states of the system, which have been the subjects of strong controversies, are argued in the ferromagnetic and antiferromagnetic phases. We show that among various magnetic structures ever proposed, only our non-collinear ones can reproduce all experimental results accumulated up to now.

## 1. Introduction

In Co oxides, the spin state is one of key elements for the understandings of their physical properties. To clarify how the spin state depends on the material parameters of the systems, various experimental studies have been carried out intensively on the oxygen deficient perovskite RBaCo$_2$O$_{5+\delta}$ (R=rare earth elements), for example, which have the linkages of CoO$_5$ pyramids and/or CoO$_6$ octahedra.[1-8] Transport and magnetic studies of the perovskite system Pr$_{1-x}$A$_x$CoO$_3$ (A=Ba, Sr and Ca)[9, 10] have also been carried out.

TbBaCo$_2$O$_{5.5}$ has the alternating stacks of CoO$_6$ octahedra and CoO$_5$ square pyramids along the *b*-axis, as shown in Fig. 1(a), and exhibits several transitions with decreasing temperature *T*, a metal to insulator transition at 340 K, to a ferromagnetic (FM) phase at 280 K and to an antiferromagnetic (AFM) one at 260 K. For this system, the non-collinear structures shown in Fig. 1(b) have been reported at *T*=270 K (in the FM phase) and *T*=250 K (in the AFM phase) by the magnetic structure analyses on single crystal samples.[5] The Co-moments in the CoO$_5$ pyramids are within the *ab*-plane and cant by the angle ~45º from the *a*-axis (or *b*-axis), while the Co$^{3+}$ ions in the CoO$_6$ octahedra are in the low-spin {LS; $(t_{2g})^6$} state, and those in the CoO$_5$ pyramids are possibly in the intermediate-spin {IS; $(t_{2g})^5(e_g)^1$} state at both temperatures. In NdBaCo$_2$O$_{5.5}$, Co$^{3+}$ ions in CoO$_6$ octahedra are predominantly in the IS state and those in CoO$_5$ pyramids are in the high-spin {HS; $(t_{2g})^4(e_g)^2$} state.[7] These results indicate that not only the local arrangement of O ions around the Co ion sites but also the ionic radii $r_R$ of R$^{3+}$, which affect the size of the polyhedra, are important for the determination of the spin state of Co$^{3+}$. For smaller $r_R$, the LS (IS) state of Co ions is expected to become more stable in CoO$_6$ (CoO$_5$) polyhedra. These relationships can be understood by the consideration of the crystal-field strength at the Co sites.

On the other hand, magnetic structures and spin states different from those described above have been reported for RBaCo$_2$O$_{5.5}$ (R= Tb, Nd and Gd) by several groups to explain transport and magnetic properties and/or the magnetic reflections observed by the powder neutron diffraction, as summarized in Fig. 2.[11-15] However, these structures except the ones proposed by Plakhty *et al*. in ref. 14 encounter certain difficulties in explaining our neutron magnetic reflection data taken for the single crystal sample,[5] as described later, although the models in refs. 12 and 15 were proposed after the presentation of our neutron diffraction studies. Because the model in ref. 14 presents the magnetic reflections at the same $Q$ points as those of our model in the reciprocal space, we cannot easily exclude it by our previous neutron data. Therefore, we have carried out the NMR studies to see the internal magnetic fields at their crystallographically distinct Co sites, because the existence or non-existence of $Co^{3+}$ ions in the LS state, gives us decisive information to distinguish which of the models proposed by Plakhty *et al*. and by the present authors' group correctly describes the magnetic structures of the FM and AFM phases. Neutron data have also been accumulated for the further confirmation of the results. Based on the analyses of these data, we show in the present paper that among various magnetic structures ever proposed, only the non-collinear structures can reproduce all experimental results accumulated up to now in a consistent way.

The present paper is composed as follows. In **2**, the experimental details are shown. In 3.1, the NMR data at 5 K, which indicate the existence of the LS $Co^{3+}$ sites, are presented. After showing the neutron diffraction data taken for single crystals of $TbBaCo_2O_{5.5}$ and $NdBaCo_2O_{5.5}$ in 3.2, we show in 3.3 the existence of the LS $Co^{3+}$ sites even at the temperature region, where the magnetic structure analyses were carried out by neutron diffraction studies. Conclusions are given in **4**.

## 2. Experiments

Single crystals of $TbBaCo_2O_{5.5}$ and $NdBaCo_2O_{5.5}$ were grown by a floating zone (FZ) method as was described in the previous reports.[5,7] The obtained crystals were annealed to control the oxygen number. Polycrystalline samples of $YBaCo_2O_{5.5}$ were prepared by a solid-state reaction.[4] These samples were confirmed by powder X-ray measurements not to have an appreciable amount of impurity phases.

The NMR measurements with nonzero applied field ***H*** were carried out by a standard coherent pulsed NMR method for the single crystal of $TbBaCo_2O_{5.5}$ with the condition of ***H***//*c*. The spectra were taken by recording the spin echo intensity with the applied field being changed stepwise. The zero-field NMR measurements in the magnetically ordered phase were also carried out, where the frequency was changed stepwise. For polycrystalline sample of $YBaCo_2O_{5.5}$, similar measurements were also carried out in the same experimental conditions as those of $TbBaCo_2O_{5.5}$ for the intensity-comparison.

In the discussion of neutron results, the data obtained in our previous works are also considered. All the neutron measurements were carried out for the single crystals of $TbBaCo_2O_{5.5}$ and $NdBaCo_2O_{5.5}$ by using the triple axis spectrometer TAS-2 installed at JRR-3 of JAEA in Tokai. The details of the experimental conditions were described in refs. 5 and 7. In the analyses, the space group P*mmm* is adopted, and the unit cell size is described by $\sim a_p \times 2a_p \times 2a_p$,[16,17] where $a_p$ is the lattice parameter of the cubic perovskite cell. Because the ***a***\*- and ***b***\*-domains exist in the used crystal, both (0,*k*,*l*) and (*h*,0,*l*) points in the reciprocal space can be observed. Several 0*kl* reflections can be distinguished from *h*0*l* reflections by the difference between the lattice parameters *a* and *b*/2 (The parameter *a* is slightly smaller than *b*/2.).

## 3. Experimental Results and Discussion
3.1 NMR Studies

NMR studies of $^{59}$Co have been carried out in the AFM phase for both the polycrystalline sample of YBaCo$_2$O$_{5.5}$ and a single crystal of TbBaCo$_2$O$_{5.5}$ to observe the internal magnetic fields at the crystallographically distinct Co sites. The zero-field NMR spectra and the field swept NMR spectra taken at 5 K are shown in Figs. 3 and 4, respectively. In these figures, the data of YBaCo$_2$O$_{5.5}$ reported by Itoh *et al.*[18] are also shown. They are roughly scaled to the present data. In our experiments, the NMR frequency $f$ was 45 MHz. (In the experiments by Itoh *et al.*, the frequency $f$ was 40 MHz.) For these spectra, the corrections of the transverse relaxation rates $T_2$ were made. In the measurements of the two samples of YBaCo$_2$O$_{5.5}$ and TbBaCo$_2$O$_{5.5}$, the experimental conditions were carefully kept to be identical. In the figures, the signal intensities of YBaCo$_2$O$_{5.5}$ and TbBaCo$_2$O$_{5.5}$ are so scaled as to show the values for equal molar numbers. From the data, the integrated intensities of the NMR signal of TbBaCo$_2$O$_{5.5}$ are found to be roughly equal to those of YBaCo$_2$O$_{5.5}$ observed by the zero-field NMR. It should be noted, however, that YBaCo$_2$O$_{5.5}$ and TbBaCo$_2$O$_{5.5}$ have the polycrystalline and single crystalline forms, respectively. Because the temperature of the AFM ordering of the Tb-moments is ~3.5 K in TbBaCo$_2$O$_{5.5}$,[19] the static internal field created at the Co sites by the Tb-moments is zero at the present experimental temperature (~5 K). However, the transverse relaxation time $T_2$ of TbBaCo$_2$O$_{5.5}$ is much shorter (~1/10) than that of YBaCo$_2$O$_{5.5}$ due to the effect of the strong fluctuations of the Tb-moments.

Two sets of $^{59}$Co signals have been found for the single crystal of TbBaCo$_2$O$_{5.5}$. The first set found by the zero-field NMR measurements has a splitting due to the *eqQ* interaction with the quadrupole frequency $\nu_Q$ ~ 35 MHz, as shown in Fig. 3(b). The NMR frequency of the central line is ~215 MHz, which corresponds to the large internal magnetic field of ~ 21.5 T at one of two distinct Co sites, say Co1 sites.

The second set of signals has been observed in the field swept NMR measurements in the region around $H$~ 4 T, as shown in Fig. 4(b) for $H$//c. The peak is observed only at the field lower than that of the free $^{59}$Co line by ~ 0.5 T. It indicates that the internal field at Co sites distinct from Co1 sites (Co2 sites) is within the *c*-plane. If the internal field is not within the plane, the signal should be found at both sides of that of free $^{59}$Co. The estimated value of the internal field is ~ 2 T. From the peak width (~ 1 T), the quadrupole frequency $\nu_Q$ is estimated to be less than ~5 MHz. Assuming that both Co1 and Co2 sites have the ordered magnetic moments and the same isotropic hyperfine coupling constants, we find that the magnetic moment at Co2 site is 1/10 of that of Co1 site.

The existence of the Co sites with large and very small magnetic moments has been found in YBaCo$_2$O$_{5.5}$ and EuBaCo$_2$O$_{5.52}$, too. However, in these Y- and Eu-based systems, peaks have also been observed at 19 MHz and 24 MHz, and at 39 MHz and 62 MHz, respectively,[18,20] indicating that there exist various Co sites with small but different internal fields in these systems.

In the present measurements, the signal has not been detected in the frequency region between 15 MHz and 80 MHz in the zero-field NMR measurements, indicating that the Co sites can be divided into only two sites, with the large and small internal magnetic fields. It is consistent with the magnetic structure model with two different magnetic moments, one is large and another one is very small or negligibly small, as discussed later.

The results of the present NMR study suggest that Co1 is in the IS or HS state, while Co2 is primarily in the LS state. The small shift of the Co2 line from that of free $^{59}$Co nuclei may be considered to originate from the transferred hyperfine field.[18,20] Because the $\nu_Q$ value of the Co1 sites is large (~35 MHz), the sites are in the pyramid, while the Co2 sites with $\nu_Q \leq 5$ MHz are in the octahedra. That is, the $Co^{3+}$ ions in the $CoO_5$ pyramids are in the IS or HS state, and those in the $CoO_6$ octahedra are in the LS state, which is consistent with the results of $YBaCo_2O_{5.5}$ and $EuBaCo_2O_{5.5}$.[18,20]

The magnetic structure by Plakhty *et al.*, which does not have the LS state, seems to contradict the present NMR results. Further considerations to extract the firm conclusion on this point are given after the results of neutron studies are presented.

3.2 Neutron Diffraction Studies

To definitely clarify which one of the models proposed by other groups and by the present authors' group is appropriate to the present systems, we have also carried out neutron diffraction studies and added new data to our previously published ones. The peak intensities of several Bragg reflections observed by neutron scattering on the single crystal of $TbBaCo_2O_{5.5}$ are summarized in Fig. 5, where the data except those shown in the top panel were reported in our previous paper.[5] With decreasing $T$ from the $T$ region above 300 K, additional reflection components, which correspond to the FM ordering, appear at $T_C \sim 280$ K. With further decreasing $T$, the components, suddenly disappear at $T_N \sim 260$ K. In our previous paper, we showed that these additional components are magnetic.[5] We also showed by determining the accurate scattering angles of several reflections, as shown in the insets of Fig. 5,[5] that the 1/201, 020 and 010 (magnetic) reflections have significant intensities, while the intensities of the 011, 100 and 1/200 ones are zero or negligibly small. These results clearly show that the magnetic reflections observed in the FM phase have not only the contribution from the ferromagnetically ordered component but also from the antiferromagnetically ordered one. For example, the 1/201 reflection is considered from the AFM component, while the 020 reflection is from the FM one. Therefore, in the FM phase, the system has the complex magnetic structure with both FM and AFM components. It has been found to be true for $NdBaCo_2O_{5.5}$, too,[7] suggesting that this type of complex magnetic structure is common to $RBaCo_2O_{5.5}$ with various species of R.

The magnetic components observed in the ferromagnetic phase disappear at $T_N \sim 260$ K with decreasing $T$, indicating that the rather abrupt change of the magnetic structure occurs. The magnetic structure obtained just below $T_N$[5] is re-drawn in Fig. 1(b). The period of the moment system along $c$ is $\sim 4a_p$ (see the bottom panel of Fig. 5, too). We will present comparative discussion later on the magnetic structures in the FM and AFM phases of $TbBaCo_2O_{5.5}$ with those proposed by other groups, to clarify which is the most probable one.

In the following arguments on the magnetic structures, we assume that the R-ion dependence can be essentially neglected. It is supported by the following facts: Similar successive transitions, the metal to insulator transition and those to the FM and then to the AFM phases are commonly observed for various species of R with decreasing $T$. Moreover, various similar behaviors such as the characteristic negative magnetoresistance[3,11,12] and metamagnetic behavior[5,11,12], for example, can be observed in association with the successive transitions, indicating that the similar magnetic states are realized by the transitions. Although the R-ion size is known to often affect the magnitude of the Co moments, it is not important for the present arguments on the magnetic structure, because as we have pointed out above, similar types of magnetic structure are realized in the systems with $Nd^{3+}$ and $Tb^{3+}$, the largest and the smallest ions among the presently relevant ions $Nd^{3+}$, $Gd^{3+}$ and $Tb^{3+}$.

Taskin *et al.* have proposed the magnetic structures shown in Fig. 2(a) on the basis of their transport and magnetization measurements using detwinned single crystals of GdBaCo$_2$O$_{5.5}$.[11,12] (In the AFM phase, besides the structure shown in the right panel of Fig. 2(a), they proposed similar possible structures with the 4$a_p$ period along the *c*-axis, and with the 4$a_p$ period along both the *b*- and *c*-axes, too.) However, the existence of the 1/2 0 1 magnetic reflection, for example, observed in the FM phase (see the top panel of Fig. 5) cannot be explained by the model shown in the left panel of Fig. 2(a). Among three possible magnetic structures they proposed in the AFM phase, the structures with the 4$a_p$ period along *b* and with the 4$a_p$ period along both *b*- and *c*-axes are incorrect, because the superlattice reflections which indicate the existence of the 4$a_p$ period along *b* have not been observed. The remaining one with the 4$a_p$ period along *c* is found not to reproduce the magnetic intensities observed in the previous studies[5] at all.

Fauth *et al.* have proposed the magnetic structures shown in Fig. 2(b) by the powder neutron diffraction studies on a polycrystalline sample of NdBaCo$_2$O$_{5.47}$[13] Their AFM and spin-state ordered (SSO) phases correspond to the FM and AFM ones of the present paper, respectively. (Fauth *et al.* did not observe the ferromagnetic component in their powder neutron diffraction studies.) However, the magnetic structure shown in the left part of Fig. 2(b) cannot explain the *T*-dependence of the observed 020 magnetic reflection of TbBaCo$_2$O$_{5.5}$ shown in the middle panel of Fig. 5, because this reflection indicates the existence of the ferromagnetic component. (The intensity of this 020 magnetic reflection is consistent with the magnetization measurement.) The existence of the magnetic reflections due to the ferromagnetic component of the ordered moment has also been confirmed for NdBaCo$_2$O$_{5.5}$, too.[7] The ferromagnetic component exists even in zero magnetic field. It should be added that the 010 reflection observed by our measurements[5] does not exist for the Fauth's model. Therefore, the model is incorrect in their AFM phase (or in the present FM phase).

In the low temperature phase (their SSO phase or the present AFM phase), the magnetic reflections expected by the Fauth's model are at the same ***Q***-positions as those by the present authors' model and the intensities of the magnetic reflections calculated by Fauth's model are roughly equal to those of our model, except the ones with weak intensities. Therefore, in the present neutron diffraction studies, we cannot distinguish if Fauth's model for the low temperature phase is correct or not.

Plakhty *et al.* have proposed the magnetic structures shown in Fig. 2(c) by powder neutron diffraction studies on polycrystalline samples of TbBaCo$_2$O$_{5+x}$ (*x*~0.5), which exhibits three phase transitions with decreasing *T*, to the ferrimagnetic phase at 290 K, to the higher-*T* AFM (AFM1) phase at 255 K and to the lower-*T* AFM (AFM2) phase at 170 K.[14] (Their ferrimagnetic phase corresponds to the FM one of the present paper.) In all these phases, the system has two distinct Co sites for the CoO$_5$ pyramids and two Co sites for the CoO$_6$ octahedra. All these sites carry large magnetic moments ( >1.5 μ$_B$) at low temperatures.

For the FM phase, Plakhty *et al.* adopted the ferrimagnetic structure different from our non-collinear one to explain both the FM and AFM components. For their AFM1 phase, Plakhty *et al.* adopted the magnetic structure ferrimagnetic within the *c* plane. In both the FM and AFM1 phases, the ***Q***-positions, where the magnetic reflections appear in their models, are same as those of our models. Therefore, we cannot distinguish if Plakhty's models are correct or not by the existence or non-existence of the magnetic reflections. However, the magnetic structure by Plakhty *et al.*, which does not have the LS state, is considered to contradict the NMR results, as discussed in 3.3 in detail.

Khalyavin has proposed the magnetic structures for the FM and AFM phases by arguing the symmetry of the system.[15)] It is difficult to distinguish if the models are correct or not from the positions of the observed magnetic reflections, because the models predict the magnetic reflections at the same positions as those of two sets of models proposed by our group and Plakhty *et al.* However, his model can be excluded because they can reproduce none of the neutron data observed by our group and Plakhty *et al.*

The Plakhty's and Khalyavin's models have the structure with the space group P*mma*, where two distinct Co sites exist for each group of the $CoO_5$ and $CoO_6$ polyhedra, that is, there are four different values of the Co moments.[21)] The unit-cell size is $\sim 2a_p \times 2a_p \times 2a_p$. In contrast to this model, we choose, as in the previous work, the space group P*mmm* with the unit cell size of $\sim a_p \times 2a_p \times 2a_p$, which has only one site for each group of $CoO_5$ and $CoO_6$ polyhedra.[5,16,17)] It is rationalized by the following facts. We have checked by neutron diffraction experiments on the single crystals of $TbBaCo_2O_{5.5}$ and $NdBaCo_2O_{5.5}$ if the nuclear superlattice reflections expected for P*mma* really exist or not. The results are negative as shown in Fig. 6, for example: For the distorted structure of Plakhty's model,[14)] the nuclear reflection peak as large as that shown by the dashed line in the figure should be observed at (1/2,1,1) point in the reciprocal space (referring to the unit cell size of $\sim a_p \times 2a_p \times 2a_p$). However, the data do not exhibit such a peak at all. The powder neutron diffraction study by Plakhty *et al.* have not been detected any evidence for the distortion, either. Thus, there is not the firm basis to consider the four different values of the Co moments adopted by Plakhty *et al.* and Khalyavin.

Now, we can summarize the discussion on the neutron results by noting followings. The models proposed by Taskin *et al.* and Fauth *et al.* can be excluded by the existence or non-existence of the magnetic reflections at certain *Q*-points. The superlattice reflections caused by the structural distortions, which the Plakhty's and Khalyavin's models are based on, have not been detected in the crystals of $TbBaCo_2O_{5.5}$ and $NdBaCo_2O_{5.5}$, or the reflection is, at least, much smaller than those expected by the X-ray data reported for $GdBaCo_2O_{5.47}$. This result indicates that there is no particular reason to adopt their complicated structural models.

3.3 Comparative Discussion between the Results of NMR and Neutron Measurements

As discussed above, only the model proposed by Plakhty *et al.* cannot be ruled out by the magnetic reflection data. However, we have clarified by the NMR studies that the LS sites of $Co^{3+}$ ions exist at 5 K, whereas the Plakhty's model does not have the LS sites in the present AFM and FM phases. Now, only the remaining problem is if the LS sites exist at the temperatures, where the magnetic structure analyses were carried out ($\sim 100$ K$<T<280$ K). The information can be found in the $T$ dependence of the intensities of the Bragg reflections from the AFM component: As shown in the bottom panel of Fig. 5 for the 001/2 reflection, for example, there does not seem to exist the significant $T$ dependence in the intensities below 100 K. It indicates that the magnetic state does not undergo the significant change below 100 K. Therefore, the magnetic structure proposed at 100 K (the AFM2 phase) by Plakhty *et al.*, which does not have the LS state, is considered to contradict the present NMR results.

On the other hand, to see if the LS state found by the present NMR studies persists up to 250 K, where our magnetic structure analyses were carried out, we examine the detailed $T$ dependence of the accumulated data of the neutron measurements up to 250 K. First, we note that there exist anomalies at 100 K and 250 K in the $T$ dependence of the 020 and 100 reflection intensities (see the middle panel of Fig. 5). These anomalies can be considered to be due to structural transitions, as described in ref. 5: If these reflections are magnetic, the spontaneous magnetization should exist, because the anomalies appear at the ferromagnetic reflection points in the reciprocal space. We can expect that effects of these structural transitions on the magnetic structure analyses made at low temperatures are very small, because the intensities of the magnetic reflections are insensitive to the small lattice distortions. Therefore, we analyze the neutron data at 7 K and 110 K by ignoring the weak superlattice peaks observed at $Q=(0,k',l'\pm 1/4)$ with even $k'$ and odd $l'$ below 100 K.[5]

As the results of the analyses, essentially the same magnetic structures shown in the right panel of Fig. 1(b) have been obtained at all the temperatures of 7, 110 and 250 K, although the canted angle and magnitude of Co-moments depend on $T$. The optimized values of the magnetic moments in the $CoO_5$ pyramids are shown in Fig. 7 and those in the $CoO_6$ octahedra are smaller than 0.03 $\mu_B$, indicating that the $Co^{3+}$ ions have the LS state in the $CoO_6$ octahedra and the IS state in the $CoO_5$ pyramids. Each $CoO_2$ layer has the canted ferromagnetic structure similar to the structure of 270 K shown in the left panel in Fig. 1(b). The obtained values of the canted angle of Co-moments in the pyramids are also shown in Fig. 7, where the values in the ferromagnetic phase ($T$= 270 K) are also shown. The canted angles are measured from the $a$- or $b$-axis. At low temperatures, the magnitudes and canted angles of the Co-moments in the pyramids approach ~2 $\mu_B$ and ~45°, respectively. For the realization of the canted magnetic structures, the long-range exchange couplings must be introduced. Although the value of ~45° may have certain significance, we have not identified the origin of the characteristic structure. These points remain as the subjects of the future studies.

## 4. Conclusions

We have carried out the NMR studies on a single crystal sample of $TbBaCo_2O_{5.5}$. New data of neutron diffraction are added to our previously published data. By analyzing both the NMR spectra and diffraction data, the magnetic structures and the spin states of the system, which have been the subjects of strong controversies are clarified in the ferromagnetic phase at temperature $T$= 270 K and in the antiferromagnetic phase at $T$= 250 K. Based on the analyses of these data, we have shown that among various magnetic structures ever proposed, only our non-collinear ones can explain all experimental results accumulated up to now.

Acknowledgments - Work at the JRR-3 was performed within the frame of the JAEA Collaborative Research Program on Neutron Scattering. The work is supported by Grants-in-Aid for Scientific Research from the Japan Society for the Promotion of Science (JSPS) and by Grants-in-Aid on priority areas from the Ministry of Education, Culture, Sports, Science and Technology.

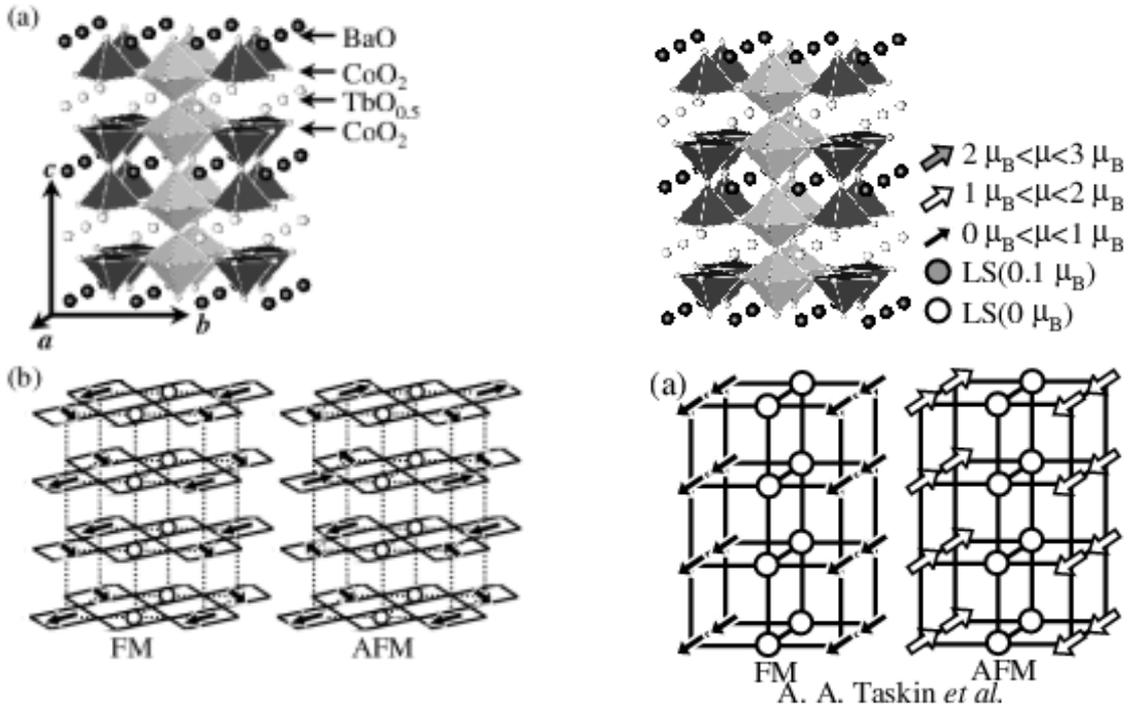

Fig. 1 (a) Oxygen deficient perovskite structure of $TbBaCo_2O_{5.5}$ is shown schematically. (b) Magnetic structures of $TbBaCo_2O_{5.5}$ reported previously by the present authors' group for the FM (left) and AFM (right) phases. The magnetic moments are shown by the arrows at the Co sites corresponding to those in (a). The open circles indicate the LS state. They are in the $CoO_6$ octahedra.

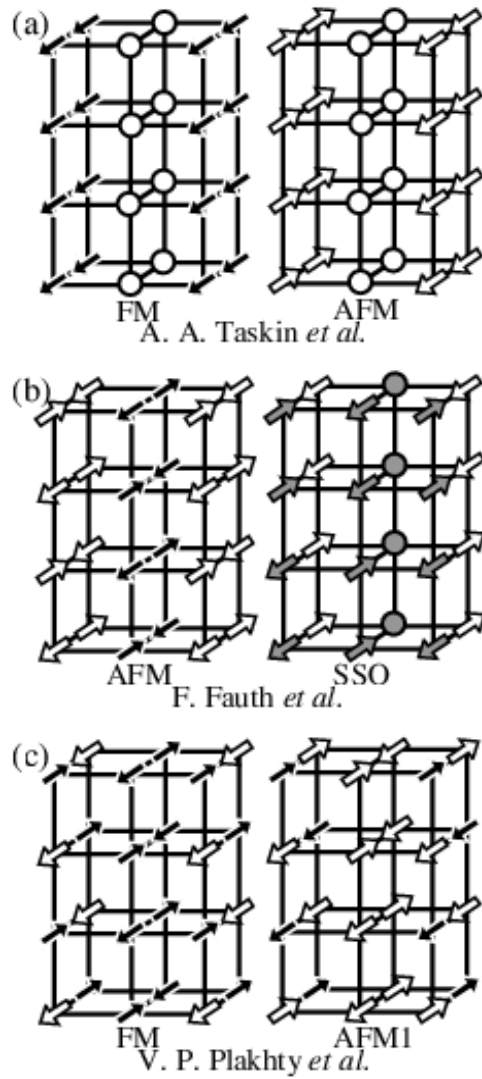

Fig. 2 Magnetic structure of $RBaCo_2O_{5.5}$ suggested by several groups are shown schematically. The magnetic moments are shown by the black, white and gray arrows at the Co sites corresponding to those shown in the top figure. These three kinds of arrows indicate the magnitudes of the Co-moments of $0\ \mu_B < \mu < 1\ \mu_B$, $1\ \mu_B < \mu < 2\ \mu_B$ and $2\ \mu_B < \mu < 3\ \mu_B$, respectively. The white and gray

circles indicate the low spin states of $Co^{3+}$ ions, having 0 $\mu_B$ and 0.1 $\mu_B$, respectively. (a) Magnetic structures of $GdBaCo_2O_{5.5}$ proposed by Taskin *et al.* are shown schematically in the FM (left) and the AFM (right) phases. (b) Magnetic structures of $NdBaCo_2O_{5.47}$ proposed by Fauth *et al.* are shown schematically in their AFM phase (the FM phase in the present paper) and their spin-state ordered (SSO) phase (the AFM phase in the present paper). (c) Magnetic structures of $TbBaCo_2O_{5.5}$ proposed by Plakhty *et al.* are shown schematically in the FM (left) and the AFM1 (right) phases.

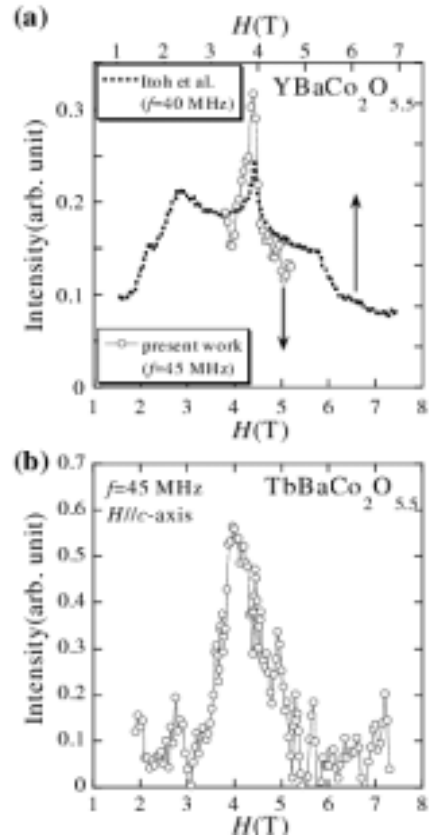

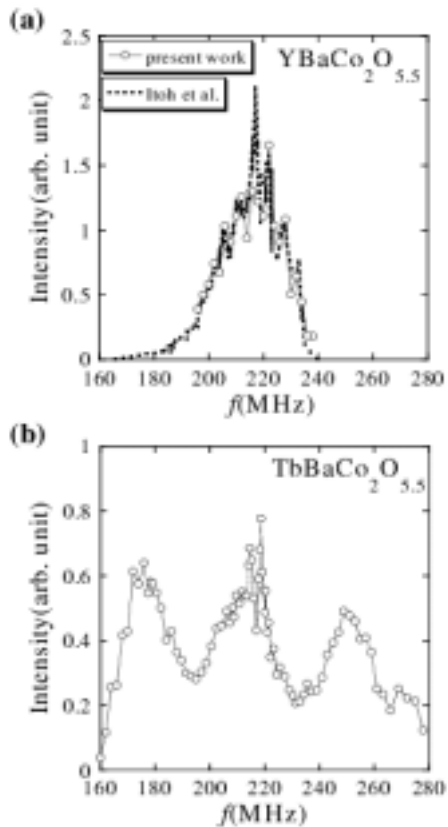

Fig. 3  Zero-field NMR spectra taken at 5 K for $TbBaCo_2O_{5.5}$ and $YBaCo_2O_{5.5}$. Open circles show the spin-echo intensities obtained by the present work. The dashed line shows the spectra reported by Itoh *et al.* [18]

Fig. 4  Field swept NMR spectra taken at 5 K for $TbBaCo_2O_{5.5}$ and $YBaCo_2O_{5.5}$. Open circles show the spin-echo intensities obtained by the present work with the NMR frequency $f$ = 45 MHz. The dashed line shows the spectra reported by Itoh *et al.*[18] with the NMR frequency $f$ = 40 MHz. The abscissas are scaled by considering the difference between the $f$-values of the measurements.

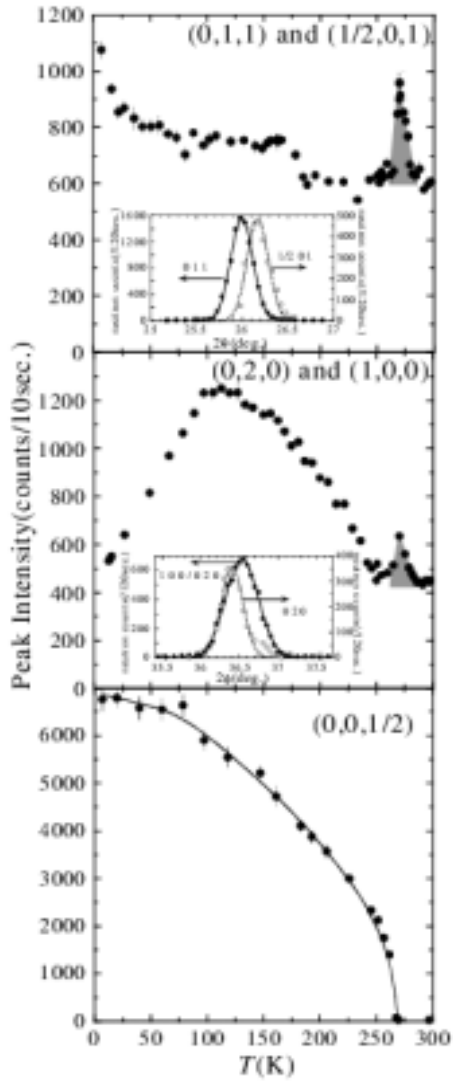

Fig. 5 Peak intensities of several reflections are shown against *T*. The shaded regions indicate the additional components of magnetic reflections in the ferromagnetic phase. Inset shows the profiles of the 011 or 1/201 (top panel) and the 020 or 100/020 (middle panel) reflections. Solid circles represent the reflection intensities at 300 K and the open circles show the magnetic scattering intensities at 270 K, which are extracted by subtracting the neutron count number at 300 K from the data at 270 K at each scattering angle 2θ.

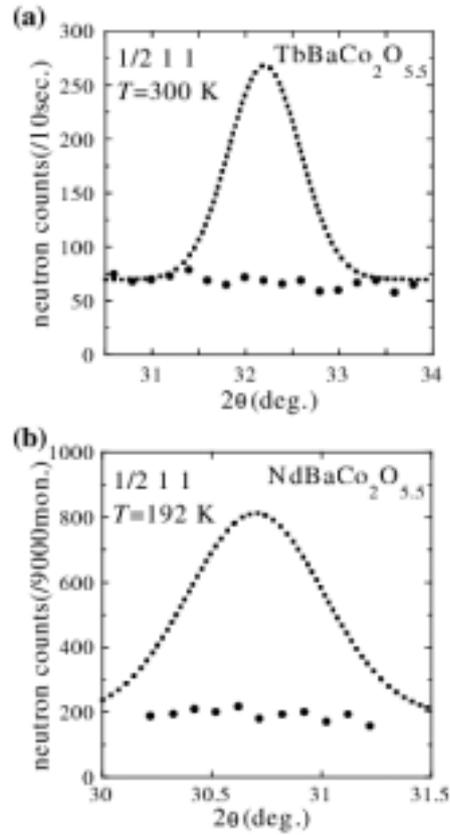

Fig. 6 Neutron intensities around the point $Q$=(1/2,1,1) in the reciprocal space are shown. The dashed line in (a) shows the intensities estimated for the present sample of TbBaCo$_2$O$_{5.5}$ by the distortion reported in ref. 14 at 300 K. (b) The similar figure to (a) is shown for the present sample of NdBaCo$_2$O$_{5.5}$ at 192 K.

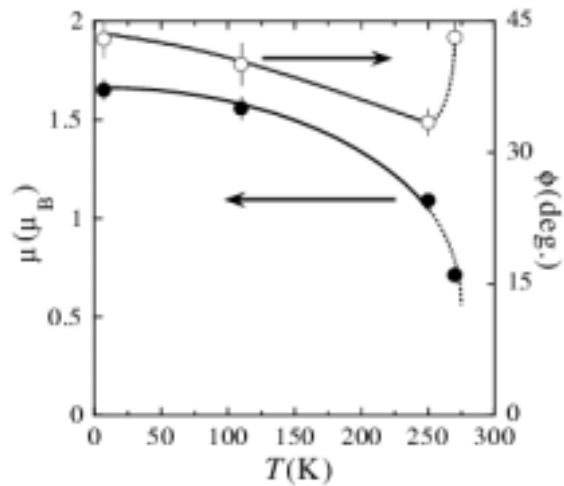

Fig. 7 Magnitude μ (solid circles) and the canted angle φ (open circles) of the Co-moments in the CoO$_5$ pyramids obtained by using the magnetic structures shown in Fig. 1(b).